\documentclass{article}
\usepackage{psfig}
\usepackage{graphicx}
\title{Quantifying the information transmitted in a single stimulus}
\author{
Michele Bezzi\\
Accenture Technology Labs, Sophia Antipolis, France\\
\texttt{michele.bezzi\ @accenture.com} \\
}
\begin{document}

\maketitle

\begin{abstract}
Shannon mutual information provides a measure of
how much information is, on average,  contained in a set of neural activities about a set of stimuli. 
It has been extensively used to study neural coding
in different brain areas. To apply a similar approach to investigate single stimulus encoding, we need to introduce a  quantity specific for a single stimulus.
This quantity has been defined in literature by four different measures, but none of them 
satisfies the same intuitive properties (non-negativity, additivity), that
characterize mutual information. 
We present here a detailed analysis of the different meanings and properties of these four  definitions.  
We show that all these measures satisfy, at least, a weaker additivity condition, i.e. limited to the response set.  This allows us to use them for analysing correlated coding, as we illustrate in a toy-example from hippocampal place cells.
\end{abstract}

\section{Information theory and mutual information}

Information theory~\cite{Shannon49} provides a natural mathematical 
framework to answer the question: {\it how much} information is contained 
in the neural patterns.
Usually, in an experiment, we choose a controlled sample of stimuli 
${\cal S}$, and we record the elicited neural responses 
$r \in {\cal R}$ when one stimulus $s \in {\cal S}$ is 
repeatedly presented with a known a priori probability $p(s)$.
%For example $r$ can be represented as the number of spikes recorded in a fixed time
%window after stimulus presentation, or the inter-spike intervals, or the exact
%timing of the first spike. 
From these data, we can estimate the corresponding 
joint probabilities $p(r,s)$ and the probability 
distribution of responses averaged over the stimuli $p(r)$;
and then   compute 
the mutual information\footnote{Estimating  joint probabilities
from experimental data   needs very large samples, 
and it is often unfeasible. Various approximate methods have been proposed to overcome this issue, see~\cite{Treves95,Strong98}.}:

 \begin{equation}
I({\cal S};{\cal R}) = \sum _{s \in {\cal S}, r \in {\cal R} }
  p(r, s) \ \log _2 \left[\frac{p(r, s)}{p(r) p(s)}
\right]=\sum _{s \in {\cal S},{r \in {\cal R}}}
 p(s) p(r|s) \ \log _2 \left[\frac{p(r|s)}{p(r)}
\right],
\label{Info1}
\end{equation}
(with  conditional 
probability $p(r|s)= p(r, s)/p(s)$ according to Bayes' rule)
or, equivalently, introducing the entropy of a probability distribution:
$H({\cal R})=-\sum_{r \in {\cal R}} p(r) \log_2 p(r)$, 
 \begin{equation}
I({\cal S};{\cal R}) = H({\cal R}) -H({\cal R}|{\cal S})= \sum_{s \in {\cal S}} p(s) [H({\cal R})- H({\cal R}|s)]
\label{Info2}
\end{equation}
where $H({\cal R}|{ s}) \equiv \sum_{s \in {\cal S}}
p(s) H({\cal R}|{\cal S})$ is the conditional entropy.

Mutual information summarizes the average amount of knowledge  we gain about the stimulus by observing neural responses (or vice-versa); e.g.: in the trivial case, they are completely uncorrelated, $p(r, s)= p(r) p(s)$
and $I=0$.
 
Mutual information has some
mathematical properties that agree to our intuitive notion of information.
In particular, we expect that  any  observation  does not decrease the knowledge we have about  the system.
So, mutual information has to be positive, as it can be easily shown 
starting from Shannon definition.
Furthermore, if we observe the response
from two different neurons or two different aspects of single unit response, $\{{\cal R}^{1},{\cal R}^{2}\}$,
the overall information   about the stimulus set 
$I(\{{\cal R}^{1},{\cal R}^{2}\};{\cal S})$
is the sum of the information contained in the first response ${\cal R}^{1}$,
plus the information we gain by reading the second response ${\cal R}^{2}$ given 
we know ${\cal R}^{1}$ ({\it chain rule}), i.e.:

\begin{equation}
I(\{{\cal R}^{1},{\cal R}^{2}\};{\cal S})=I({\cal R}^{1};{\cal S})+
I({\cal R}^{2};{\cal S}|{\cal R}^{1})
\label{chainR}
\end{equation}
Relationship~(\ref{chainR}) may be used to investigate the independence
of coding in cell population or in presence of multiple response features~\cite{Schneidman04}.
For example, if  $\{{\cal R}^{1},{\cal R}^{2}\}$ 
 encode different features of the stimulus independently\footnote{Note, here we are dealing with information independence.
This is different a concept from the independence
between the two different stimulus set (i.e. $p(x,d)=p(x)p(d)$) and also from conditional independence
(i.e. $p(x,d|r)=p(x|r)p(d|r)$). These are three distinct, although interconnected, measures of independence (for a detailed discussion
of the topic see~\cite{Schneidman04}).}, then the information 
they convey about the stimulus  
has to be the sum of the information conveys separately, i.e.
$I(\{{\cal R}^1,{\cal R}^2\},{\cal S}\})=I({\cal R}^1,{\cal S})+
I({\cal R}^2,{\cal S})$.
 Thus, by evaluating this information (or synergy function~\cite{Schneidman04}) we can quantitatively estimate the different
contributions to the stimulus encoding from the different response features. Note that this {\it additivity} property directly relies on the validity of the chain rule~(\ref{chainR}).
Similarly, due to its symmetric form, we can express the chain rule in terms
of two different stimuli features (e.g. color and shape in visual 
stimuli) $\{{\cal S}^{1},{\cal S}^{2}\}$
\begin{equation}
I({\cal R};\{{\cal S}^{1},{\cal S}^{2}\})=I({\cal R};{\cal S}^{1})+
I({\cal R};{\cal S}^{2}|{\cal S}^{1})
\label{chainS}
\end{equation}

\section{Stimulus specific information}

In many cases it may be interesting to know which  stimulus
contained more information in a given set~\cite{BezziNiArDaCo04, Machens05}, 
or investigate how a single stimulus is encoded in terms of different response features.
In his original formulation, Shannon did not provide any insights about
how much information can be carried by a single symbol, such as a single
stimulus in our case. 
After Shannon's seminal work, many definitions of one-symbol specific
information have been introduced, usually referred as {\it stimulus specific information}
in the neural processing context.
% Since all of them have been proposed in 
%the framework of neural response analysis, and in particular for investigating
%information carried by a single stimulus, they are usually referred as 
%{\it stimulus specific information}.
Ideally,  
{\it stimulus specific information} should be  proper information in a 
mathematical sense (non-negative, additive) and give 
mutual information when averaged over the stimulus
set.
%Unfortunately there is no possible definition of stimulus specific 
%information  that fulfills these properties, but four different 
%definitions have been proposed up to now
Four different  alternative definitions have been proposed  in literature. We give here  a detailed analysis of their features and 
 different meanings\footnote{We deal with
possible definitions of the {\it same quantity}, so all these definitions
are usually referred to as ``stimulus specific information''. For sake of simplicity, we extend the notation of~\cite{DeWeese99} to all four definitions, referring to them as $I_1$, $I_2$, $I_3$ and $I_4$. We  also mention their {\it original} names as introduced in~\cite{Bezzi02,Butts03}.}.
In particular, we investigate the different roles of stimulus and response regarding additivity rules, Eqs.(\ref{chainR},~\ref{chainS}), (Section~\ref{additivityR}) and show a possible application (Section~\ref{Hippo}). 
%as well as
%presenting a
%novel relationship  between surprise and local information (Section ...).
In Table~\ref{Table:Properties} we summarize the main features of these four quantities.

We conclude that there is  no ``fully'' satisfactory 
definition of this quantity, in the sense that no one of these
definitions shares all mathematical properties
of Shannon mutual information, which
is so appealing from the application point of view;  but each of them can be used to 
investigate  different aspects of single stimulus information transmission.

\subsubsection*{Stimulus specific surprise}

Originally proposed by Fano~\cite{Fano}, this  definition can be immediately inferred from
Eq.~(\ref{Info1}), simply taking the single stimulus contribution to the sum:
\[
I_1(s)=Surprise(s)=\sum_{r \in {\cal R}}p(r|s) \log_2 \frac{p(r|s)}{p(r)} 
\]
This quantity measures the deviation (also called Kullback-Leibler distance)
between the marginal distribution $p(r)$ and conditional probability distribution
$p(r|s)$.  It clearly averages to the mutual information,
 i.e.  $\sum_{s \in {\cal S}} p(s) I_{1}(s)=I({\cal S};{\cal R})$, 
and it is always non-negative: $I_1(s) \geq 0$ for
 $s \in {\cal S}$.
Furthermore it is the only positive decomposition of the mutual information
(for the proof, see Appendix 2 in Ref.~\cite{DeWeese99}).
Since $I_1(s)$ is large when $p(r|s)$ dominates in the regions where 
 $p(r)$ is small, i.e. in presence of {\it surprising} events, this quantity is often referred as
 ``stimulus specific surprise'' or simply ``surprise''.
 Surprise lacks 
additivity, and this causes many difficulties when we want to apply it to a
sequence of observations. Despite this main drawback, specific surprise
has been widely used in neural coding literature (see for example~\cite{Rolls97,Lu04,Paz04}).
  
\subsubsection*{Stimulus specific information}  

An entropy based  definition has been proposed by De Weese and Meister~\cite{DeWeese99}  and it  may be derived from
Eq.~(\ref{Info2}),  extracting the single stimulus contribution from the sum:
\[
I_2(s)= H({\cal R})-H({\cal R}|s)=
-\left[\sum_{r \in {\cal R}}p(r) \log_2 p(r) -p(r|s) \log_2 p(r|s)\right] 
\]

Here, information is identified with the reduction of entropy between marginal distribution $p(r)$
and conditional probability $p(r|s)$.
This quantity captures  the reliability of the neural response for 
a given stimulus. Indeed, 
it expresses the difference of uncertainty between the a priori 
knowledge of the response set, $H({\cal R})$, and after stimulus
presentation $H({\cal R}|s)$. 
A stimulus characterized by highly predictable responses
has a large $I_2(s)$ value. This means that we can easily predict
a response when we know the stimulus, but not necessarily
vice-versa.
 
 As shown in~\cite{DeWeese99}, this is  the only decomposition of mutual information
  that is also additive, but, unlike mutual information, it can assume negative values. 

Note that any weighted combination of
$I_1$ and $I_2$ averages to mutual information, and it can represent a possible definition
of stimulus specific information.
Thus, we have an  infinite number of plausible choices for a
stimulus-dependendent decomposition of mutual information. 
But, as mentioned above,  only $I_1$ is always non-negative and for $I_2$ only the chain rule
\ref{chainS} is fulfilled.  
 
\subsubsection*{Stimulus specific information (SSI) (2)}  
 More recently Butts~\cite{Butts03} introduced a new definition, which emphasizes the causal relationship between stimulus and response in neural processing:
 
\begin{eqnarray*}
	I_3(s)= SSI(s) &=&  \sum_{r \in {\cal R}} p(r|s) I_2(r) =\sum_{r \in {\cal R}}p(r|s) \left[ H({\cal S})-H({\cal S}|r) \right] \\
	 &=& H({\cal S})- \sum_{r \in {\cal R}} p(r|s)H({\cal S}|r)	
\end{eqnarray*}
where $I_2(r)$ is one-symbol specific information  applied to a single neural response:
 $I_2(r)=H({\cal S})-H({\cal S}|r)$. 
This measure represents the average reduction of uncertainty (difference of entropy)
 after an observation of the response 
given the stimulus, or, in other words, the stimulus-conditioned average of the 
response-specific information $I_2(r)$. 
So, if one stimulus is characterised by a very informative response, in the sense 
of $I_2(r)$ (such as responses that almost completely determine the stimulus), it results 
in a large value of $I_3$.

\subsubsection*{Stimulus information density $I_l$} 

A fourth definition has been  proposed 
by Bezzi {\it et al.}~\cite{Bezzi02}
in the framework of position-encoding in hippocampal formation.
\begin{eqnarray}
	I_4(s)= I_l(s) &=&  I({\cal R};\{s,\bar{s}\})=  \\
	 &=& \sum_{r \in {\cal R}} 
p(s) p(r| s) \log_2\left[ \frac{p(r | s)}{p(r)}
\right]  +  p(\bar{s}) p(r |\bar{s} ) \log_2\left[ \frac{p(r|\bar{s})}
{p(r)} \right] ,
\label{Il}
\end{eqnarray}
where $\bar{s}$ is the complementary set of $s$,
i.e. $\bar{s} \equiv \bigcup_{s' \in {\cal S}, s' \neq s}s'$
(with $s \bigcup \bar{s} ={\cal S}$).
 In other words, we partition the set of stimuli 
into two subsets,
one containing  the stimulus  $s$ only and the complementary set
composed by all the other stimuli, then we compute the average mutual information using 
these two stimuli set.
This is a measure of how well we can distinguish between the single stimulus $s$ and 
all the others observing the neural response ${\cal R}$. 
$I_4$ is an actual mutual information, thus it is  
positive and additive (this last condition holds only for a very particular choice of the stimuli, see below),
but it does not average to  mutual information of the whole
stimulus set
(i.e. $\sum_{s \in {\cal S}} p(s) I_4(s) \neq I({\cal S};{\cal R})$). 

This quantity has a simple relationship with specific surprise $I_2$
for very unlikely stimuli $p(s) \ll 1$.
Indeed, expanding Eq.~(\ref{Il}) 	for $p(s) \ll 1$ we get:
\begin{eqnarray}
	I_l(s) &=&    
p(s) \sum_{r \in {\cal R}} p(r|s) \log_2\left[ \frac{p(r | s)}{p(r)}
\right]  +  \mathcal{O}({p(s)^2})
\label{loc-surp}
\end{eqnarray}
that corresponds to $p(s) I_1(s)$.
This expression has two consequences: it supplies an additional 
theoretical justification to specific surprise 
and assures that if we have a large set of
stimuli characterized by $p(s) \ll 1$ for each ${s \in {\cal S}}$
the sum of the local information converges to the average mutual information.

\begin{table}[tbf]
\begin{center}
\begin{tabular}{||c|c|c|c|c||} \hline
Definition & Positive  &  Chain rule  & Chain rule   & Average MI \\
 &  definite &  (Responses) &  (Stimuli) &  \\  \hline
$I_1$ Surprise & Yes & Yes & No & Yes  \\ \hline 
$I_2$ & No & Yes & Yes & Yes \\ \hline 
$I_3$ & No & Yes &  No & Yes \\  \hline
$I_4$ & Yes & Yes  & Yes/No & No \\ \hline 
\end{tabular}
\end{center}
\caption{Main properties of the four definitions of stimulus specific information.} \label{Table:Properties}
\end{table}

\section{A weaker additivity condition}
\label{additivityR}
Among the above four definitions  for stimulus specific information, only  $I_2$ is  additive (obeying Eqs.~(\ref{chainR},~\ref{chainS}))~\cite{DeWeese99}.
But, despite not being additive in a rigorous sense,
 the other three quantities
still fulfill relationship~(\ref{chainR})\footnote{Due to space limitation, we do not report the full proofs here. In short, the proof for $I_1$ follows from additivity property for entropy, for $I_2$ and $I_3$ applying twice Bayes' rule to $p(r_1,r_2|s)$ and using the normalisation condition for $p(r|s)$, $I_4$ is an actual mutual information so additive.}, but not~(\ref{chainS}), or in other words, they are additive limited to 
the response set. 
This results in a weaker additivity property, which may still be used for investigating
how different features of the response encode a single stimulus, such as testing independence (see Section~\ref{Hippo} for an example). 
We should remark that due to the different meanings of the four definitions,
this analysis  can give contradictory results depending on the measure chosen.
A special case is represented by $I_4$, because 
it is a proper mutual information (with a particular choice of stimulus set),
then we expect it to be additive in terms of stimuli, too. But, $I_4(s)$ requires an additional constraint: the {\it two} subsets partition, $\{s,\bar{s}\}$, of stimulus set ${\cal S}$. This last condition cannot be preserved in general for the chain rule, Eq.~(\ref{chainS})~\cite{Bezzi02}.

\section{Stimulus specific information in hippocampal place cells}
\label{Hippo}

\begin{figure}[t]
\begin{center}
 \includegraphics[totalheight=10.cm]{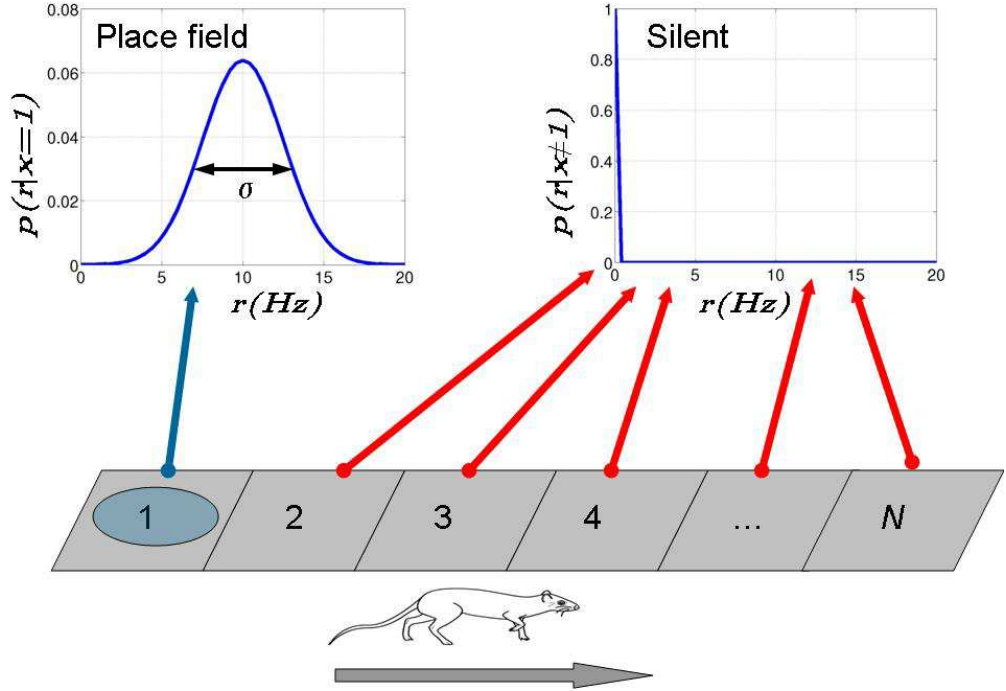}
\end{center}
\caption{A rat running in  linear track (unidirectional). Position in the track is binned, 
in $N$ spatial bins.  Conditional probability
distribution for a place cell: $x_1$, place field (top, left) and $x_{i\neq 1}$ (top, right), outside place field.}
\label{rat}
\end{figure}

\begin{figure}[t]
\begin{center}
\includegraphics[totalheight=6.cm]{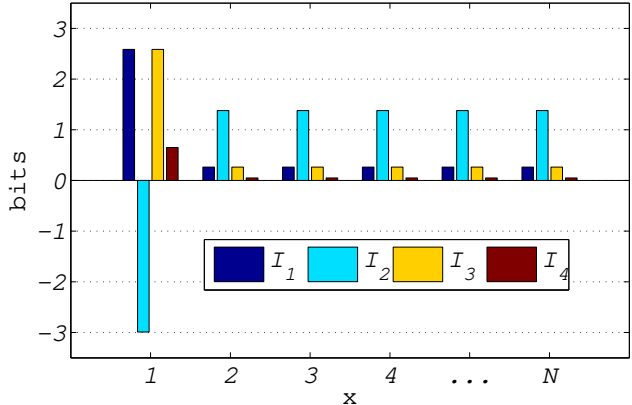}
\end{center}
\caption{Numerical values of stimulus specific information for  different locations $x$. Left, $I_{1,2,3,4}$ considering only firing rate. Place field corresponds to $x=1$. Parameter values:  $N=6$, $\sigma=2.5 Hz$ and $\overline{r}=20 Hz$ for the place field profile.}
\label{infos}
\end{figure}

\begin{figure}[t]
\begin{center}
 \includegraphics[totalheight=10.cm]{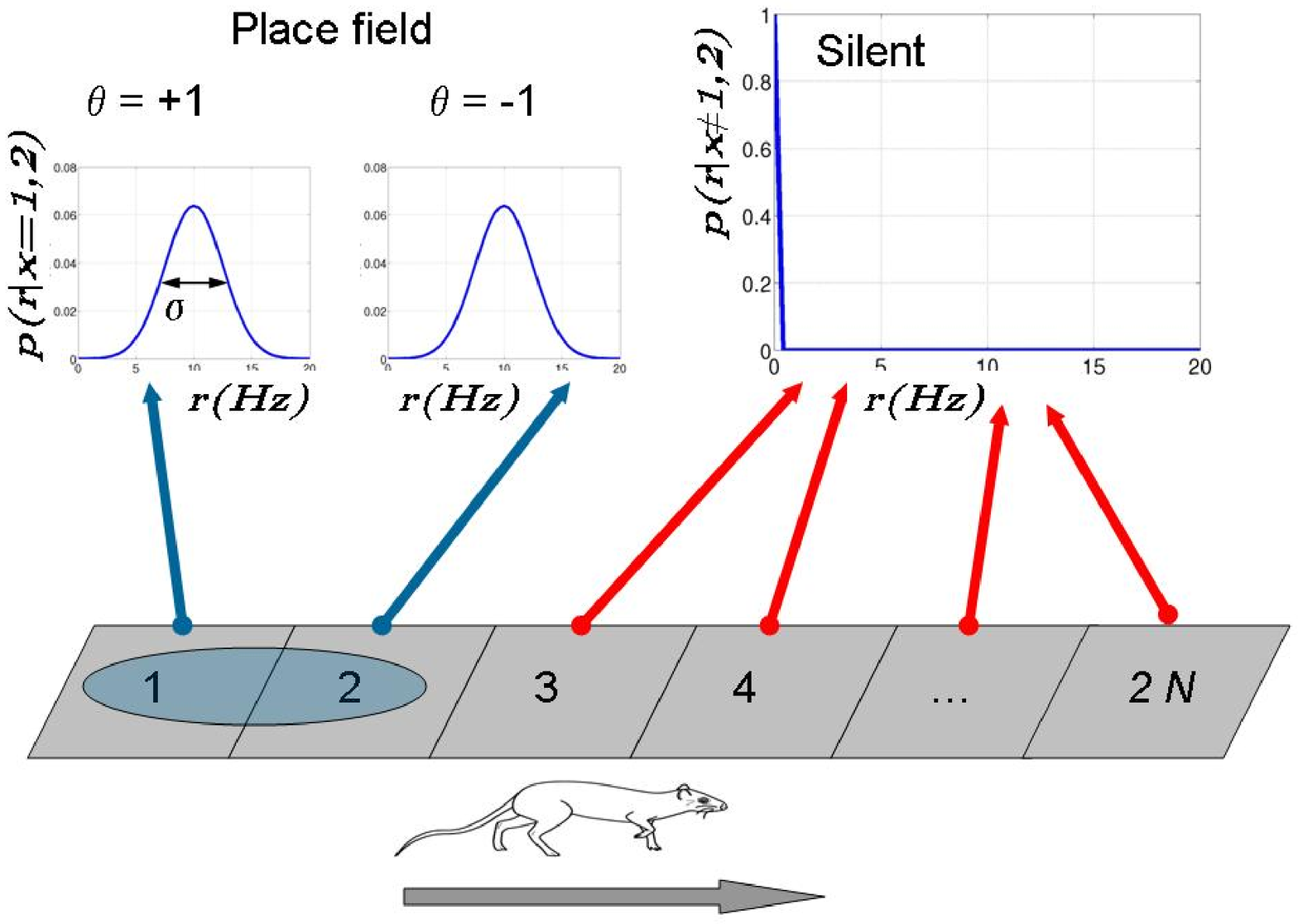}
\end{center}
\caption{A rat running in  linear track (unidirectional). Position in the track is binned, 
in $2 N$ spatial bins.  Conditional probability
distribution for a place cell: $x_1$,$x_2$ place field (top, left) for $\theta=1,2$, respectively, and $x_{i\neq 1,2}$ (top, right), outside place field.}
\label{rat2}
\end{figure}

\begin{figure}[t]
\begin{center}
\includegraphics[totalheight=6.cm]{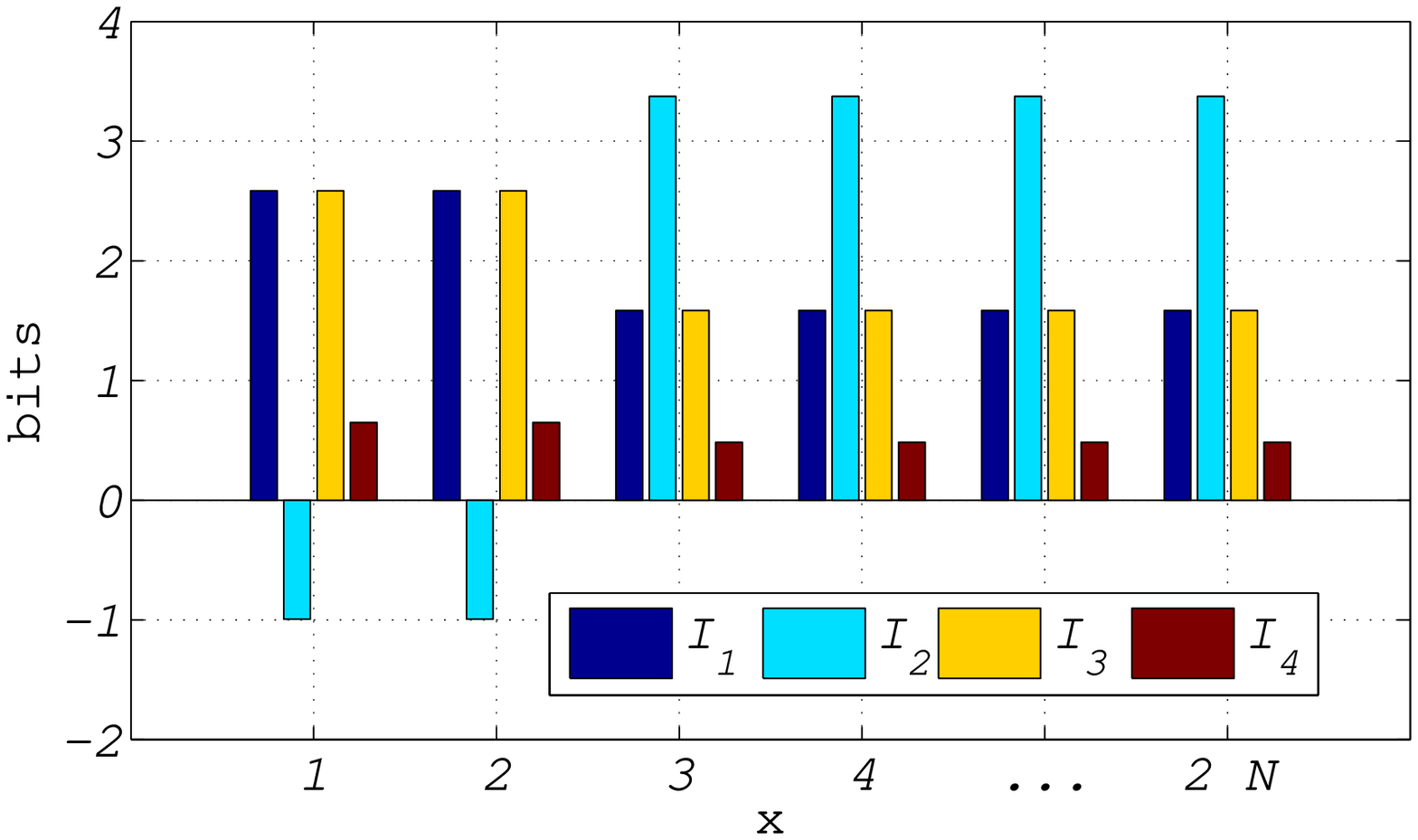}
\end{center}
\caption{Numerical values of stimulus specific information for  different locations $x$. $I_{1,2,3,4}$ considering  firing rate and phase-shift (binary values). Place field corresponds to $x=1$ and $x=2$. Parameter values:  $N=6$, $\sigma=2.5 Hz$ and $\overline{r}=20 Hz$ for the place field profile.}
\label{infos2}
\end{figure}

To show the different behavior of stimulus specific informations in a more 
realistic setting, we consider here the example of spatial encoding in hippocampal place cell. 

Place cells  selectively fire at an elevated rate when the animal is in a
particular location of an environment and, in some cases (e.g. linear tracks),
moving in one specific direction.
For sake of simplicity we do not consider direction here.
Place cell firing profile is then characterized by a silent part
(when the animal is not in the place field), and an active part
corresponding to the place field. 
Let us consider an ideal experiment in which a rat is moving 
in a linear track in one direction at constant speed (from left to right in Fig.~\ref{rat}). 
The set of  locations is the stimuli set, that we consider discrete
and finite, composed by $N$ elements. We indicate with ${x \in {\cal X}}$ a single position with $p(x)=1/N$. $x_1$ corresponds to 
the place field, and we measure the  firing rate of the cell $r$,
 in each spatial bin $x$. 
Assuming a gaussian profile inside the place field (see Fig.~\ref{rat}, top-left), and considering the cell completely silent elsewhere (i.e. neglecting fluctuations, Fig.~\ref{rat}, top-right), we are able to compute analytically $I_{1,2,3,4}$ and the mutual information. Results are summarised in Fig.~\ref{infos}.
$I_2(x_1)$ may be negative if $\sigma$ is large enough, 
due to the fact that the uncertainty of the response
$H({\cal R}|x)$  could be larger than average uncertainty in the correspondence to the place field.
This implies that place-field corresponds to the less informative
stimulus according to $I_2$ measure.
On the contrary, using  the other definitions  $x_1$ is largely 
the most informative location in agreement 
with our intuitive notion that the cell codes for the position corresponding
to the place field.  

To better illustrating the role of additivity rule (Eq.~\ref{chainR}), let us consider a simple extension of the previous example, where two different aspects of single cell neural response are considered.
In a similar setup, we measure the average firing rate and 
the phase shift respect to a specific rhythm: the theta oscillations $\Theta$.
Indeed, during locomotion hippocampus is characterized by a high-amplitude 4-8 Hz oscillation,
called theta rhythm. Theta oscillations play an important role in spatial 
encoding.  Cell spiking  shifts gradually to earlier phases of the theta cycle as the animal 
moves through the cell place field (theta phase precession)\cite{OKeefe93}.

For sake of simplicity, in our toy-example, we take this quantity to assume just two values: $\theta=+$ 
(say a phase shift of $270^o$) and $\theta=-$ (say a phase shift of $90^o$), furthermore
we  associate  this feature to non place field location (this is not in general appropriate for 
real place cells), see Fig.~\ref{rat2}. So each cell response is represented by two quantities:
firing rate and average phase shift, where the latter is a binary variable.
We take finer spatial bins, in this way the place field is covered by two spatial bins.
Place cell fires in the place field, corresponding to positions 
$x_1$ and $x_2$, with a gaussian profile (as shown in Fig.~\ref{rat2} (top-left)). 
For $x_1$, spike times follow on average theta oscillatory maximum of $\theta=+$
and, vice versa, in $x_2$ they come earlier ($\theta=-$). Similarly in the other
odd positions $\theta=+$, and  $\theta=-$ in the other even positions, where
the cell is  silent. We neglect fluctuations and we assume that firing rate is zero in these locations.

Overall we have $2 N$ different positions. Assuming, as before, all the positions are
equally likely $p(x) = \frac{1}{2N}$ and $p(\theta=+)=p(\theta=-)=1/2$, we can analytically estimate 
mutual information and all the stimulus specific information.
Results are summarised in Fig.~\ref{infos2}.
As in the previous example, place field location corresponds to most informative stimulus according to $I_{1,3,4}$, but less informative for $I_2$. Furthermore, in this example, we can test if $r$ and $\theta$ encode different aspects of stimulus $x$ independently or not, checking if:
$I(\{{\cal R},{\Theta}\}, x)=I({\Theta},x)+I({\cal R}, x)$.
These terms can be computed analytically using the assumptions above. 
We find that, using $I_{1,2,3}$ phase and average firing rate 
encode a single position $x$ independentely. Since  averaging these quantities over the stimulus set we have mutual information, independence condition is satisfied for  mutual information.
On the contrary we have redundant encoding using $I_4$.  
This discrepancy between the behavior of $I_{1,2,3}$ and $I_4$ (and mutual information too) is due to the fact that, considering  local information,  we implicitly correlate the two response features putting together all the locations different from $x_1$,
so that phase and average firing rate are mixed together in $p(\{{\cal R},{\Theta}\}| \bar{x_1})$
 
Note, to  apply the same criterium to test encoding independence for different stimuli features $S_1, S_2$ (e.g. location and direction) only $I_2$ may be used.

\section{Conclusions}

Shannon mutual information is symmetric for stimulus and response set. When we consider stimulus specific information,  we lose this symmetry property, resulting in a weaker additivity condition, limited to the response set. 
So, although none of definitions of stimulus specific information has all the mathematical properties of mutual information, all of them may be used for exploring correlated encoding by multiple response features.
In addition, we have shown that different stimulus specific information measures are suitable to answer different questions, and, not surprisingly, they may assume different numerical values in the same context. In conclusion, there is no unique definition, but the {\it right} measure should be chosen according to the aspect of neural coding we are interested in studying.

\subsubsection*{Acknowledgments}
We thank  A. Arleo and F. Linaker for critical reading of the manuscript.

\end{document}